\documentclass[secnumarabic,amssymb, amsmath,12pt, 
nofootinbib,tightenlines, nobibnotes, aps, prl]{revtex4}
\usepackage{graphicx}
\begin{document}
\newcommand{\be}{\begin{equation}}
\newcommand{\ee}{\end{equation}}
\newcommand{\ba}{\begin{eqnarray}}
\newcommand{\ea}{\end{eqnarray}}
\newcommand{\bea}{\begin{eqnarray*}}
\newcommand{\eea}{\end{eqnarray*}}
\newcommand{\ts}{\textstyle}

\bigskip
\vspace{2cm}
\title{Structure-dependent radiative corrections  to $\phi \to 
K^+K^-/K_LK_S$ decays}
\vskip 6ex
\author{F. V. Flores-Ba\'ez$^1$}
\author{G. L\'opez Castro$^2$\footnote{On sabbatical leave from: 
Departamento de F\'\i sica, Cinvestav, A.P. 14-740, 07000 M\'exico D.F.}}
\affiliation{$^1$Departamento de F\'{\i}sica, Cinvestav,  
Apartado Postal 14-740, 07000 M\'exico D.F., M\'exico \\
$^2$Instituto de F\'{\i}sica, Universidad Nacional Aut\'onoma de M\'exico, 
04510 M\'exico D.F, M\'exico}
\bigskip

\bigskip

\bigskip

\begin{abstract}
Current predictions for the ratio of $\phi \to  K^+K^-/K_LK_S$ decay 
rates exceed the corresponding experimental value in about five standard 
deviations. By far, the dominant sources of isospin breaking to this 
ratio are the phase-space (52$\%$) and the electromagnetic radiative 
(4.3$\%$, computed within scalar QED) corrections. Here we estimate the 
effects of the electromagnetic structure of kaons and other 
model-dependent contributions into the radiative corrections.
 \end{abstract}

\maketitle
\bigskip

\section{Introduction}

Precise knowledge of the ratio for $P^+P^-/P^0\overline{P}^0$ production 
($P=K,\ D$ or a $B$ pseudoscalar meson) is very important for 
measurements of branching fractions and determination of fundamental 
parameters at the 
$\phi,\ \psi(3770)$ and $\Upsilon(4S)$ resonances \cite{ratio}. 
 It has long been known that standard theoretical calculations 
overestimate the isospin breaking corrections to the ratio  of $\phi \to 
K^+K^-/K_LK_S$ decay rates \cite{Bramon:2000qe}. Thus, the isospin 
breaking  corrections induced by the mass difference of kaons and the 
electromagnetic radiative corrections, change the ratio
\be
R  \equiv \frac{\Gamma(\phi\to K^+K^-)}{\Gamma(\phi\to K^0\overline{K}^0)} 
\ , 
\ee
from unity to about 1.59 \cite{Bramon:2000qe}. This value lies 
4.7$\sigma$'s 
above the corresponding experimental value \cite{Yao:2006px}:
\be
R_{exp}= 1.45 \pm 0.03\ ,
\ee
quoted by the PDG from an {\it overall fit} to measured branching 
fractions of $\phi$ decays.  This large discrepancy is still missing a 
convincing 
explanation. 

 Large isospin breaking corrections to the  $\phi \to 
K^+K^-/K^0\overline{K}^0$ ratio are naturally 
expected since both decay modes occurs near threshold in a  $p$-wave. 
However, it is  rather difficult to identify additional contributions 
(beyond conventional phase-space and radiative corrections) which 
may render theory 
and experiment into agreement. For instance, calculations of the isospin 
breaking to the ratio of $\phi K^+K^-$ and  $\phi K^0\overline{K}^0$ 
coupling constants, carried out in the context of effective hadronic  
interactions, may increase further the theoretical prediction up to 
$R^{theory}=1.62$ 
\cite{Bramon:2000qe,Benayoun:2001qz}. Related to this problem, the effects 
of strong scattering phases on the isospin breaking effects to 
$P^+P^-/P^0\overline{P}^0$ ($P=K,\ D$ or $B$ meson states) production in 
$e^+e^-$ annihilation near 
threshold where considered in Ref. \cite{Dubynskiy:2007xw}.  
Finally, we should mention that a non conventional mechanism to solve this 
discrepancy which proposes corrections to the  Fermi Golden rule formula 
of decay rates was discussed in ref. \cite{Fischbach:2001ie}. 

   In this paper we revisit the calculation of radiative corrections to 
the $\phi \to K^+K^-,\ K^0\overline{K}^0$ decays. The corrections of  
order $\alpha$ to the decay into charged kaons were calculated long ago by 
Cremmer and Gourdin \cite{Cremmer:1969er} using scalar QED. Since  
the electromagnetic structure of kaons is ignored in the framework 
of scalar QED, no corrections to the neutral mode are induced in this 
case. In the present paper we focus on the electromagnetic form factors 
of kaons and compute their effects in the observable $R$ defined in eq. 
(1). One should note that this calculation was first  considered in 
\cite{Bramon:2000qe}, where a correction of order $2\times 
10^{-3}$ was found for the decay rate into charged kaons. In 
that paper, the correction to the neutral mode is mention to be 
negligible. We find good agreement with the findings of ref. 
\cite{Bramon:2000qe} and provide and independent test of the model by 
considering that the dominant contributions come from the region of 
validity of the vector meson dominance model. In addition, we
also consider the effects of the sub-leading contributions due to 
hard-photon emission. Since structure-dependent effects have been 
found to be important in other calculations of radiative corrections, for 
instance in  $\tau \to  \pi\nu_{\tau}$ and $\pi \to \mu\nu_{\mu}$ decays 
\cite{Decker:1994ea}, it is worth  seeing if they can be important in 
the $\phi$ decays of our interest. 

\section{Isospin breaking corrections to $R$ in scalar QED}
  
  We start by defining the tree-level amplitude for the 
$\phi(q,\eta) \to K(p)\overline{K}(p')$ decays: 
\be
{\cal M}_0(K\overline{K}) =ig^j(p-p')\cdot \eta,\ \ 
\ee
where $g^+$ ($g^0$) refers to the strong coupling constant for the 
$K^+K^-$ ($K^0\overline{K}^0$) final state, and $\eta^{\mu}$ is the 
polarization  four-vector of  the $\phi$ vector meson ($q\cdot \eta=0$). 
The corresponding ratio of decay rates reads: 
\be
R^{theory}_0= \left(\frac{g^+}{g^0} \right)^2\cdot \frac{v_+^3}{v_0^3} \ . 
\ee
where $v_+$ ($v_0$) denotes the velocity of the charged (neutral) kaons in 
the  rest frame of $\phi$ meson. In the limit of isospin symmetry, 
$g^+=g^0$  and 
$m_{K^+}=m_{K^0}$, thus the ratio of decay rates becomes $R^{theory}_0=1$. 

  The largest source of isospin breaking corrections to $R^{theory}_0$  
arise from the mass difference of charged and neutral kaons 
[$(v_+/v_0)^3=1.5225$]. The next most important breaking effect comes from 
radiative corrections. The corrections of $O(\alpha)$ to the   
$\phi \to K^+K^-$ decay rate, $\delta_{QED}= 
2\delta^v_{point}+\delta^r$,  were calculated in Ref. 
\cite{Cremmer:1969er} in 
the context of scalar QED. They include the sum of virtual corrections 
($\delta^v_{point}$) to the non-radiative amplitude and the real 
($\delta^r$) photonic corrections.  The 
virtual corrections are divergent for infrared photons but are finite in 
the ultraviolet region owing to the Ward identity satisfied by the vertex 
and  self-energy corrections. 
  The real photon corrections $\delta^r$ contains an infrared divergent 
piece $\delta^r_I$ and a regular contribution $\delta^r_R$, namely 
$\delta^r=\delta^r_I+\delta^r_R$. For consistency $\delta^r_I$ is 
computed by summing over the longitudinal and 
transverse degrees of freedom of a massive photon \cite{Coester}. 
The sum of virtual and soft-photon corrections,  $2\delta^v_{point}+\delta^r_i$ 
is explicitly free from infrared divergences, as it should be. The explicit 
expressions of $\delta_{point}^v$ and $\delta_I^r$ can be found in Ref. 
\cite{baez2007}.

   The calculation of the regular term $\delta^r_R$ can be done in a 
numerical way. It can receive contributions from intermediate states 
other than $K^{\pm}$  mesons (for instance, $\phi \to K^+K^{*-} \to 
K^+K^-\gamma$). However, these model-dependent corrections are expected to 
be very small either for charged (which we found to be  $-7.1 
\times 10^{-8}$ 
for the contribution of  $K^*$ intermediate state) or 
neutral \cite{KLOE} channels because 
$\omega^{max}$ is very  small compared to the masses of other hadrons. 
Thus we obtain:  
\be
\delta^r_R= 7.96 \times 10^{-5} \ .
\ee
  
When we include the phase space corrections, $(v^+/v^-)^3$, and the 
radiative corrections of scalar QED, $\delta_{QED}=2\delta^v_{point}+ 
\delta^r_I+\delta^r_R = 0.04315$,  one gets:
\be
R^{theory}=R^{theory}_0\left(1+\frac{}{}\delta_{QED}\right)= 1.588
\ ,
\ee
which is about 5$\sigma$'s above the experimental value shown in Eq. (2).

\section{Structure-dependent effects in radiative corrections to $R$}.
   
  Measurements of the electromagnetic interactions of kaons at low 
\cite{ff_low} 
and intermediate \cite{cmd2} energies exhibit an structure which 
can be well described within a vector dominance model. As usual, we 
define the kaon electromagnetic vertex $\gamma^*(k) \to 
K(p)\overline{K}(p')$:
\be
-ieF_{K}(k^2)(p-p')_{\mu}\ ,
\ee
where $k^2=(p+p')^2$ is the squared momentum of the virtual photon.  

Following refs. \cite{cmd2,Decker:1994ea} we write the form factors
 in the vector dominance model as:
\be
F_K(k^2)= \sum_{V=\rho,\omega,\phi} \frac{g_{VK\bar{K}}}{f_V}\cdot 
\frac{m_V^2}{\hat{m}_V^2-k^2} \ ,
\ee
where $em_V^2/f_V$ denotes the photon--vector-meson couplings, $m_V$ is 
the mass of 
intermediate vector meson and $\hat{m}_V^2=m_V^2-im_V\Gamma_V 
\theta(k^2-k^2_{threshold})$ for timelike $k^2$ ($k^2_{threshold}$ is 
the square of virtual momentum that allows to generate imaginary parts 
in the one-loop corrections to the $V$ meson propagator). At $k^2=0$, the 
form 
factors are normalized to the electric charges of kaons and the following 
condition  must be  satisfied:
\ba
&& \sum_{V=\rho,\omega,\phi} \frac{g_{VK^+K^-}}{f_V} =1 \\
&& \sum_{V=\rho,\omega,\phi} \frac{g_{VK^0\overline{K}^0}}{f_V} =0 \ .
\ea
A simplifying assumption (also used in the experimental analysis 
of Ref. \cite{cmd2}) is implemented by using the  SU(3)-invariant 
Lagrangian for the $VPP'$ interaction which gives the couplings:
\ba
g_{\rho K^{+}K^{-}}&=&-g_{\rho K^{0}\bar{K^{0}}}=
\frac{1}{2}G_{VPP'}\nonumber \\ 
g_{\omega K^{+}K^{-}}&=& g_{\omega 
K^{0}\bar{K^{0}}}=\frac{\sqrt{3}}{2}G_{VPP'}\sin \theta_V  \\
g_{\phi K^{+}K^{-}}&=& g_{\phi 
K^{0}\bar{K^{0}}}=\frac{\sqrt{3}}{2}G_{VPP'}\cos \theta_V \nonumber \ ,
\ea
where $\theta_V$ is the $\omega-\phi$ mixing angle (we will use here 
$\tan\theta_V=1/\sqrt{2}$ as in Ref. \cite{cmd2}) and $G_{VPP'}$ is the 
SU(3)-invariant strong coupling constant. Solving Eqs. (9,10) with the 
constraints given in (11) leads to:
\ba
T^+_{\rho}&=&-T^0_{\rho}=\frac{1}{2}\nonumber  \\
T^+_{\omega}&=&T^0_{\omega}= \frac{f_{\phi}}{2\left[f_{\phi}+ 
\sqrt{2}f_{\omega} \right]} \nonumber\\ 
T^+_{\phi}&=&T^0_{\phi}=\frac{f_{\omega}}{\sqrt{2}\left[f_{\phi}+
\sqrt{2}f_{\omega} \right]}\ ,
\ea
where we have introduced the notation $T^{+,0}_V\equiv 
g_{VK^{+,0}K^{-,0}}/f_V$ for $V=\rho,\ \omega,\ \phi$ mesons. 
Finally, if we specify the values of the electromagnetic couplings 
($T_{\omega}=0.1790,\ T_{\phi}=0.3210$, using the measured rates of 
$\omega,\ \phi\to e^+e^-$ decays) the form factors do not  contain 
further free parameters. 

  Now, let us write the form factors by separating explicitly the point 
and structure-dependent contributions:
\ba
F_{K^+}(k^2)&=& 1 + \sum_{V=\rho,\omega, 
\phi}T^+_V \left(\frac{k^2+m_V^2-\hat{m}_V^2}{\hat{m}_V^2-k^2}\right) \ 
,\\
F_{K^0}(k^2)&=& 0+ \sum_{V=\rho,\omega,\phi}
T^0_V\left(\frac{k^2+m_V^2-\hat{m}_V^2}{\hat{m}_V^2-k^2}\right) \ .
\ea
Note that for small values of $k^2$ ($k^2 \leq k^2_{threshold}$), the 
numerator of the second term in the r.h.s of the  above equations is 
linear in $k^2$

 The Feynman diagrams contributing to the virtual corrections within 
this meson dominance model are displayed in Figure 1 for the generic $\phi 
\to K\overline{K}$ decay. The contributions of self energies 
should be added to these virtual corrections.
Since the structure-dependent piece of form factors --second term in 
the r.h.s. of Eqs. (13)-(14)-- falls linearly as $k^2$ approaches zero , 
their 
contributions to radiative corrections are free from infrared 
divergences. As in the  corresponding case of scalar QED, the result is 
also free of ultraviolet  divergences.

\begin{figure}
\includegraphics[width=12cm]{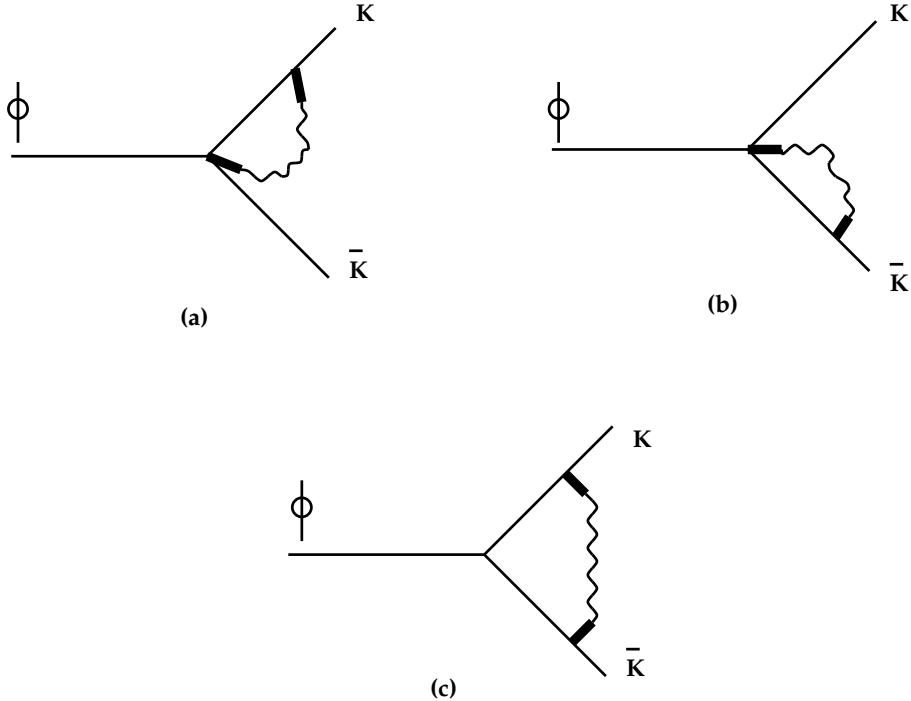}
\vspace{-0.0cm}
\caption{Feynman graphs for virtual corrections to $\phi \to 
K\overline{K}$ decays within the vector meson dominance model.}
\end{figure}

When we insert the form  factors  in the calculation of the virtual 
corrections, the expressions for the one-loop amplitudes are the same as 
in the point case but with an additional factor $|F_K(k^2)|^2$ in the 
integrand over virtual momenta. The structure-dependent pieces of the 
radiative corrected amplitudes (due to $T_V^{+,0}\not = 0$) become:
\ba
{\cal M}_{\small SD}^v ({\small K^+K^-})\!\! &=&\!\! {\cal 
M}_0({\small K^+K^-}) 
\!\left(\frac{\alpha}{4\pi}\right)\! \left\{ 
[4.37-i0.44]\!\! \frac{}{}
(T^+_{\rho})^2+[4.39+i0.04]\!\! (T^+_{\omega})^2 + 
\![5.84+i0.02]\!(T^+_{\phi})^2 
\right. 
\nonumber 
\\ && \ \ \ \ \ \ \ \ \left. + [10.25-i0.34]T^+_{\rho}T^+_{\phi} 
+ 
[9.05-i0.40]T^+_{\rho}T^+_{\omega} 
+ [10.27-i0.03]T^+_{\omega}T^+_{\phi} \right. \nonumber \\ 
&& \ \ \ \ \ \ \ \ \left. +[- 6.30+i0.53] T^+_{\rho} \frac{}{}  
+[-6.31+i0.03]T^+_{\omega}+[-7.94+i0.01]T^+_{\phi} \right\}  \\
&=&  {\cal M}_0(K^+K^-)\times \delta^+_{VMD}\ ,
\ea
and 
\ba
{\cal M}_{\small SD}^v (K^0\overline{K}^0)\!\! &=&\!\! {\cal M}_0 
(K^0\overline{K}^0)\! \left(\frac{\alpha}{4\pi} \right)\! 
\left\{ [4.34-i0.46] (T^0_{\rho})^2 \frac{}{} +[4.36+i0.02] 
(T^0_{\omega})^2 
+ [5.81+i0.01](T^0_{\phi})^2 \right. 
\nonumber 
\\ && \ \ \ \ \ \ \ \ \left. + [10.16-i0.34]T^0_{\rho}T^0_{\phi} 
+ 
[8.96-i0.40]T^0_{\rho}T^0_{\omega} \frac{}{} 
+ [10.18-i0.03]T^0_{\omega}T^0_{\phi} \right\} \\
&=&  {\cal M}_0(K^0\overline{K}^0) \times \delta^0_{VMD}\ .
\ea
where ${\cal M}_0(K\overline{K})$ denote the amplitudes at the tree-level 
defined in Eq. (3). 
In the above results, the  terms linear in $T_i^{+,0}$ appear from the 
interference between the point and the structure-dependent terms in the 
square of Eqs. (13),(14). Note that the imaginary parts arising 
from the finite decay width of vector mesons give a small 
contribution in radiative corrections. Finally, when 
we insert the numerical values for 
the couplings constants in the above expressions, we get:
\ba
\delta^+_{VMD} &=& -1.13 \times 10^{-3}\ ,\\
\delta^0_{VMD} &=& -1.37 \times 10^{-5} .
\ea
Thus, once we include the effects induced by the electromagnetic structure 
of kaons, Eq. (5) gets modified to:
\ba
R^{theory}&=& 
R_0^{theory}\left(1+\delta_{QED}\frac{}{} 
+2[\delta^+_{VMD}-\delta^0_{VMD}]\right) \nonumber \\
&=& 1.585\ .
\ea
Therefore, the structure-dependent effects in virtual radiative 
corrections are tiny but larger than hard-photon contributions (eq. 5).

 In view of the above result,  one may wonder how appropriate is using the 
meson dominance model, Eqs. (13),(14), in the full range of virtual photon 
momenta $k$. To address this 
question we introduce a modified photon propagator according to (see for 
example Ref. \cite{Decker:1994ea}):
\be
\frac{1}{k^2} \to \frac{1}{k^2}\cdot \frac{\mu^2}{\mu^2-k^2}\ .
\ee 
where $\mu$ is a cutoff scale which suppresses the contributions of high
$k^2$ values. Clearly, in the limit $\mu^2 \to \infty$ one should recover 
the previous results, while for finite values of 
$\mu^2$ the 
contributions of very high $k^2$ get suppressed.
\begin{figure}
\includegraphics[width=11cm, angle=270]{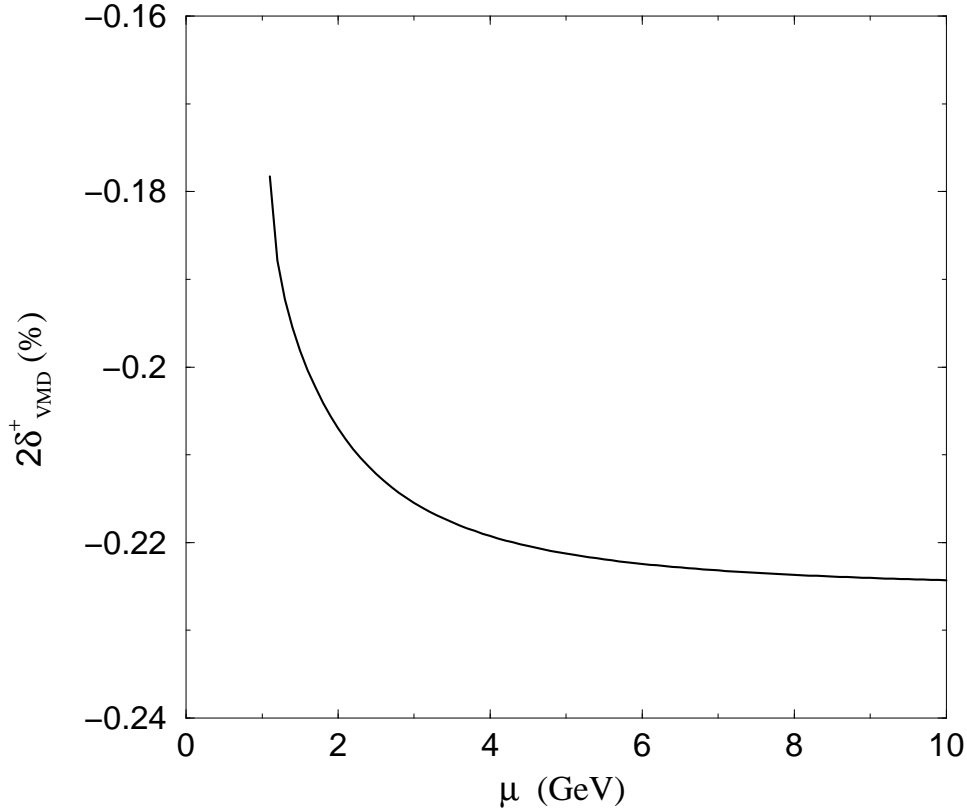}
\vspace{-0.0cm}
\caption{Structure-dependent virtual corrections $2\delta^+_{VMD}$ to the 
decay rate of $\phi \to K^+K^-$ decays as a 
function of the cutoff scale $\mu$.} 
\end{figure}

   Using in loop calculations the modified photon propagator of Eq. (22), 
we can compute again the structure-dependent parts of radiative 
corrections. Following the recipe given in Eqs. (111-113) of reference 
\cite{Decker:1994ea}, we can easily compute these corrections in the case 
of $\phi \to K^+K^-$ decays for several values of the cutoff $\mu$. Our 
results, displayed in Figure 2, show that the values of 
$\delta^+_{VMD}(\mu)$ reach the previous value already for low values of 
$\mu$, {\it 
i.e.} the most important contributions arise in the  region of validity of 
the model (intermediate energies).

\section{Summary and conclusions}

We have computed the radiative corrections to the ratio $R=\Gamma(\phi 
\to K^+K^-)/\Gamma(\phi\to K_LK_S)$ by taking into account the 
electromagnetic structure of the $K$ mesons within a vector-meson 
dominance model. Our results are in agreement with previous calculations 
of ref. \cite{Bramon:2000qe} for the charged kaon channel. Although tiny, 
the structure-dependent corrections are larger than hard real photon 
corrections. We have shown that the main contribution of 
structure-dependent corrections arise from the region of energies where 
the  vector-meson dominance of kaon form factors is expected to hold.

Clearly, structure-dependent corrections do not resolve the discrepancy 
between theory  and the experimental value of $R$ (see eq. 2 above).  It 
is worth noticing, 
however, that a weighted average of direct measurements of  $\phi \to
K\overline{K}$ decay rates gives $R_{exp}=1.49\pm 0.05$ 
\cite{Yao:2006px}, which is only 1.9$\sigma$ below the theoretical 
prediction (eq. 21). This weighted average of direct measurements is 
dominated by results of the CMD2 Collaboration \cite{cmd2}. It may happen 
that this result of direct measurements will turn out to be more reliable 
than the {\it indirect} value obtained from a constrained fit which 
requires that the 
sum of dominant decay modes saturates the total decay width of the $\phi$ 
meson \cite{note}.

 If the experimental value shown in Eq. (2) is confirmed by new 
measurements, still another possibility to solve the discrepancy  
can be provided by the short-distance effects in radiative 
corrections. Short-distance corrections can be induced by 
 highly virtual photons coupled to  $q\bar{q}$ quark pairs ($q=u,\ d,\ 
s$), in other  words, by the quark components of the 
photon wavefunction \cite{Anisovich:1996hh}. Since the light 
$q\bar{q}$ pair required to produce $K^+K^-/K^0\overline{K}^0$ in 
$\phi$ decays is $u\bar{u}/d\bar{d}$, additional isospin breaking 
correction to $R$ may be induced. In the case of radiative corrections 
to semileptonic weak decays, short-distance effects provide a 
universal 
correction (they do not break isospin symmetry) to all semileptonic 
processes \cite{sdc} since tdhey affect only the underlying quark decay 
process (isospin breaking arises only from long-distance corrections). 
 Thus, although short-distance corrections could eventually contribute 
to  isospin breaking corrections in $R$ we restrict here only to long- 
and intermediate-distance radiative corrections.

\
{\bf Acknowledgements} 

The authors acknowledge partial financial support from Conacyt.

\end{document}